\documentstyle[aps,prl,epsfig,twocolumn]{revtex}

\def\be{\begin{equation}}
\def\ee{\end{equation}}
\def\bea{\begin{eqnarray}}
\def\eea{\end{eqnarray}}
\def\bma{\begin{mathletters}}
\def\ema{\end{mathletters}}
\def\C{\hbox{$\mit I$\kern-.6em$\mit C$}}

\begin{document}
\draft

\title{Separability and entanglement in 2$\times N$ composite
quantum systems}

\author{M. Lewenstein$^1$, J. I. Cirac$^2$, and S. Karnas$^1$}

\address{$^1$Institut f\"ur Theoretische Physik, Universit\"at Hannover, 
D-30163 Hannover, Germany}
\address{$^2$Institut f\"ur Theoretische Physik, Universit\"at Innsbruck,
A-6020 Innsbruck, Austria}

\date{\today}

\maketitle

\begin{abstract}
We show  that all density operators of 2$\times N$--dimensional 
quantum systems that remain invariant after partial transposition
with respect to the first system are separable. Based on this criterion,
we derive a sufficient separability condition for general density 
operators in such quantum systems. We also give a simple proof of 
the separability criterion in $2\times 2$--dimensional systems
[A. Peres, Phys. Rev. Lett {\bf 77}, 1413 (1996)].
\end{abstract}


\narrowtext

% -------------------------------------------------------------

Perhaps, entanglement is the most intriguing property of Quantum
Mechanics. It arises when two or more systems are a
non--separable state; that is, in a state that cannot be prepared
locally \cite{We89}. Apart from having fundamental
implications\cite{Peresbook}, such as Bell's theorem or the existence of
decoherence, entanglement is in the realm of several practical
applications of Quantum Information \cite{Di95}.

Despite of its importance, we do not know yet how to quantify
entanglement. In fact, even in the simple case in which we only have two
systems $A$ and $B$ in a mixed state $\rho$, there exists no general criterion
that allows us to distinguish whether the state is separable or not;
that is, whether it can be written as 
\be
\rho=\sum_i |e_i,f_i\rangle\langle e_i,f_i|,
\ee
where $|e_i,f_i\rangle\equiv |e_i\rangle_A\otimes|f_i\rangle_B$ are
product states. An important step forward to clarify this situation was
taken by Peres \cite{Pe96} and the Horodecki family \cite{Ho96}. They
found such a criterion for the cases in which the Hilbert space
corresponding to the first subsystem is two--dimensional and the one of
the second is either two-- or three-- dimensional, the so--called
$2\times 2$ and $2\times 3$ cases, respectively. Although the criterion
is very simple, its proof involves mathematically advanced concepts
\cite{St63}, and it is not easily accessible to most physicists working
on Quantum Information. 

Let us consider the $M\times N$ case, where the Hilbert spaces $\C^M$ and
$\C^N$ corresponding to systems $A$ and $B$ have dimension $M$ and $N$,
respectively. Given an operator $X$ and an orthonormal basis
$|m\rangle_A$ ($m=1,\ldots,M$) for the first system, the partial
transpose of $X$ with respect to system $A$ in that basis is defined as
\be
\label{parttran}
X^{T_A} = \sum_{n,m=1}^M {}_A\langle n|\rho|m
\rangle_A \; |m\rangle_A\langle n|
\ee
The separability criterion introduced by Peres \cite{Pe96} states that
for $M=2$ and $N=2,3$ the density operator $\rho$ describing the state
of systems is separable iff $\rho^{T_A}\ge 0$, i.e. if its partial
transpose is a positive operator. Surprisingly enough, this criterion is
not valid for higher values of $N$ or $M$ \cite{Ho98b}. In particular, for $N=M=3$
(i.e. two three--level systems), Bennett {\it et al.} \cite{Be98} have
found families of non--separable states fulfilling $\rho^{T_A}=\rho\ge
0$. These states are related to the existence of sets of product states
that cannot be extended with other product states to form a basis
\cite{Be98}, and play a crucial role in the understanding of concepts
like non--locality \cite{Be98b} or EPR paradox without entanglement
\cite{Ho98}. One may wonder whether this class of non--separable states
exists in other dimensions. In this Letter we show that this is not the
case for $2\times N$ systems. More specifically, we show that if
$\rho^{T_A}=\rho$ then necessarily $\rho$ must be separable. Based on
this fact, we give a sufficient condition for separability in these
systems \cite{Sepcond}. Furthermore, our results also allow to derive a
proof based on simple concepts of linear algebra of the Peres criterion
in $2\times 2$ systems. 

Given the fact that some of the results we have obtained to prove our
claims may be interesting in other applications of Quantum
Information Theory, we present them in the form of lemmas. This also
leads to a more compact and readable structure of this Letter. 

We consider a density operator $\rho\ge 0$ acting on $\C^2\otimes \C^N$
and with $\rho^{T_A} \ge 0$. We will denote by $K(\rho)$, $R(\rho)$ and
$r(\rho)$ the kernel, range, and rank of $\rho$, respectively. The
partial transpose will be taken with respect to a given basis
$\{|0\rangle,|1\rangle\}\in \C^2$ of system A. We will denote by
$|e^\ast\rangle$ the complex conjugated vector of $|e\rangle$ in that
basis; that is, if $|e\rangle=\alpha|0\rangle +\beta|1\rangle$ then
$|e^\ast\rangle=\alpha^\ast|0\rangle +\beta^\ast|1\rangle$. We will use
the subscript ``r'' to denote real vectors (e.g. $|e_{\rm
r}\rangle=|e_{\rm r}^\ast\rangle$), and we will denote by $|\hat
e\rangle$ the vector orthogonal to $|e\rangle\in \C^2$, i.e. $\langle
e|\hat e\rangle=0$. Throughout this paper we will make use of the
property 
\be
\label{property}
_A\langle e_1|\rho|e_2\rangle_A 
 = _A\langle e^\ast_2|\rho^{T_A}|e_1^\ast\rangle_A 
\ee
for any pair of vectors $|e_{1,2}\rangle \in \C^2$; this property follows
directly from the definition of partial transposition (\ref{parttran}).
Unless we specify it and in order to simplify the notation, will
consider all the states of system $A$ normalized and the ones of system
$B$ unnormalized. 

In order to carry out our analysis we need to recall a lemma and a
corollary introduced in Ref.\ \cite{Le98}. For their proofs we refer the
reader to this reference.

{\bf Lemma 1:} If $|e,g\rangle\in R(\rho)$ then $\rho_1\equiv
\rho-(\langle e,g|\rho^{-1}|e,g\rangle)^{-1} |e,g\rangle\langle e,g|$ is
positive and $\rho^{-1}|e,g\rangle \in K(\rho_1)$.

{\bf Corollary 1:} Let $|e,g\rangle\in R(\rho)$, $|e^\ast,g\rangle\in
R(\rho^{T_A})$, $\lambda_1\equiv (\langle
e,g|\rho^{-1}|e,g\rangle)^{-1}$ and $\lambda_2 \equiv (\langle
e^\ast,g|(\rho^{T_a})^{-1}|e^\ast,g\rangle)^{-1}$. Then $\rho_1\equiv
\rho- \lambda |e,g\rangle\langle e,g|$ with $\lambda= {\rm
min}[\lambda_1,\lambda_2]$ fulfills $\rho_1,\rho_1^{T_A}\ge 0$ and
$r(\rho_1)=r(\rho)-1$ [$r(\rho^{T_A}_1)=r(\rho^{T_A})-1$] if
$\lambda_1\le \lambda_2$ [$\lambda_2\le \lambda_1$].
 
We will now consider some results that imply that if there is a
product vector in the kernel of $\rho$, then we can reduce the
dimensionality of the second system.

{\bf Lemma 2:} $|e,f\rangle \in K(\rho)$ iff $|e^\ast,f\rangle \in
K(\rho^{T_A})$. 

{\em Proof:} If $|e,f\rangle \in K(\rho)$ we have $0=\langle
e,f|\rho|e,f\rangle=\langle e^\ast,f|\rho^{T_A}|e^\ast,f\rangle$.
Given that $\rho^{T_A}\ge 0$ this implies that
$\rho^{T_A}|e^\ast,f\rangle=0$. Similary, we have that if
$|e^\ast,f\rangle \in K(\rho^{T_A})$ then $\rho|e,f\rangle=0$.$\Box$

{\bf Lemma 3:} If $|e,f\rangle \in K(\rho)$, then one of the following
statements is true: (i) $|\hat e,f\rangle \in K(\rho)$ and $|\hat
e^\ast,f\rangle \in K(\rho^{T_A})$; (ii) There exists a non-zero
$|g\rangle\in \C^N$
such that $\rho|\hat e,f\rangle=|\hat e,g\rangle$ and $\rho^{T_A}|\hat
e^\ast,f\rangle=|\hat e^\ast,g\rangle$.

{\em Proof:} Using Lemma 2 we have that for all $|h\rangle\in \C^N$ we
can write $0=\langle \hat e^\ast,h|\rho^{T_A}|e^\ast,f\rangle=\langle
e,h|\rho|\hat e,f\rangle$ and therefore either $\rho|\hat e,f\rangle=0$
or $\rho|\hat e,f\rangle=|\hat e,g_1\rangle$ for some $|g_1\rangle\in
\C^N$. In a similar way we have that either $\rho^{T_A}|\hat
e^\ast,f\rangle=0$ or $\rho^{T_A}|\hat e^\ast,f\rangle=|\hat
e^\ast,g_2\rangle$ for some $|g_2\rangle\in \C^N$. Using Lemma 2 we
conclude that there are two possibilities: (i) $\rho|\hat e,f\rangle=0$
and $\rho^{T_A}|\hat e^\ast,f\rangle=0$, which coincides with the first
statement of the Lemma; (ii) $\rho|\hat e,f\rangle=|\hat e,g_1\rangle$
and $\rho^{T_A}|\hat e^\ast,f\rangle=|\hat e^\ast,g_2\rangle$ for some
$|g_1\rangle,|g_2\rangle \in \C^N$. We have $|g_2\rangle=\langle \hat
e^\ast|\rho^{T_A}|\hat e^\ast,f\rangle =\langle
e|\rho|\hat e,f\rangle=|g_1\rangle$ which corresponds to the second
statement.$\Box$

{\bf Lemma 4:} If $\exists |e,f\rangle \in K(\rho)$ then we can write
$\rho=\rho_1 + \rho_s$, where $\rho_s$ is separable, $\rho_1$
is supported on $\C^2\otimes \C^{N-1}$ and $\rho_1,\rho_1^{T_A}\ge 0$.

{\em Proof:} According to Lemma 3, we can have two situations: (i)
$|e,f\rangle,|\hat e,f\rangle\in K(\rho)$ and therefore $\rho$ is
already supported on $\C^2\otimes \C^{N-1}$; (ii) $\rho^{-1}|\hat
e,g\rangle=|\hat e,f\rangle$ and $(\rho^{T_A})^{-1}|\hat
e^\ast,g\rangle=|\hat e^\ast,f\rangle$. If we define 
\be
\label{rho1}
\rho_{1}=\rho -\frac{1}{\langle \hat e,g|\rho^{-1}|\hat e,g\rangle} 
 |\hat e,g\rangle\langle \hat e,g|
        =\rho- \frac{1}{\langle g|f\rangle}|\hat e,g\rangle\langle \hat e,g|
\ee
then $\rho_{1}^{T_A}=\rho^{T_A} - (\langle \hat
e^\ast,g|(\rho^{T_A})^{-1}|\hat e^\ast,g\rangle)^{-1} |\hat
e^\ast,g\rangle\langle \hat e^\ast,g|$. According to Lemma 1,
$\rho_1,\rho_1^{T_A}\ge 0$. Moreover, $\rho_1|\hat e,f\rangle=\rho|\hat
e,f\rangle-(\langle g|f\rangle)^{-1} |\hat e,g\rangle\langle \hat
e,g|\hat e,f\rangle =0$ and $\rho_1|e,f\rangle=\rho|e,f\rangle=0$. Thus
$|e,f\rangle,|\hat e,f\rangle\in K(\rho_1)$ and therefore $\rho_1$ is
supported on $\C^2\otimes \C^{N-1}$. Note that in 
this case $r(\rho_1)=r(\rho)-1$. $\Box$

This powerful lemma simply states that if there is a product vector in
the kernel of $\rho$, then we can reduce the dimensionality of the
second system keeping positive the partial transposed operator. On the
other hand, Corolary 1 allows to reduce the rank of $\rho$, i.e. to
increase the dimension of the kernel of $\rho$, if we find appropriate
product vectors in its range. Thus, the idea is to find out whether
there are product vectors in the kernel of $\rho$. If there are, we
reduce the dimensionality. If not, then we try to find product vectors
in the range until we make the kernel ``sufficiently large'' to include
a product vector in it. Thus, the important question is to find out when
we can ensure that there are product vectors in the range or in the
kernel. The following lemma states that we can always find product
vectors in any subspace of dimension $\ge N$. 

{\bf Lemma 5:} Any subspace $H\subset \C^2\otimes \C^N$ with dim$(H)=M\ge N$ 
contains at least one product vector. 
For $M>N$ it contains at least one product
vector $|e_{\rm r},f\rangle$.

{\em Proof:} Lets us denote by $\{ |\Psi_i\rangle,
i=1,\ldots,2N-M\}$ an orthonormal basis in the orthogonal complement of $H$. We
write
\be
\label{Psi}
|\Psi_i\rangle= \sum_{k=1}^N A_{i,k}^\ast |0,k\rangle + B_{i,k}^\ast
|1,k\rangle,
\ee
and $A^\dagger A + B^\dagger B=1$. We look for product vectors in $H$
of the form $|\Phi\rangle=(\alpha|0\rangle+|1\rangle)\otimes \sum f_k
|k\rangle$, i.e.
\be
\label{AB}
(\alpha A + B)\vec f =0.
\ee
If $M>N$ we have more variables than equations, and therefore there is a
solution for all $\alpha$, and in particular for $\alpha$ real. 
For $M=N$ we have that there is always a solution since
${\rm det}(\alpha A + B)$ is a polynomial of degree $N$ in $\alpha$.$\Box$

From the results that we have obtained so far, we have the following

{\bf Theorem 1:} If $r(\rho)=N$, and $K(\rho)$ does not contain any pair
$|e,f\rangle,|\hat e, f\rangle$ [i.e. $R(\rho)$ is not supported on
$\C^2\otimes \C^{N-1}$], then $\rho$ is separable.

{\em Proof:} We use induction. For $N=1$ the statement is true. Let us
assume that it is true for $N-1$. If $r(\rho)=N$, then $K(\rho)$ has
dimension $N$, and according to Lemma 5 it contains a product vector
$|e,f\rangle$. Lemma 3 tells us that $\rho|\hat e,f\rangle=|\hat
e,g\rangle$, with $|g\rangle \ne 0$ and from Lemma 4 we obtain
$\rho=\rho_1 + \rho_s$, with $\rho_1\ge 0$ supported on $\C^2\otimes
\C^{N-1}$ and $r(\rho_1)=N-1$. Now we show that $\rho_1$ cannot be
supported on $\C^2\otimes \C^{N-2}$, which would complete the proof. If
it were, there would be two vectors $|e,h\rangle,|\hat e,h\rangle\in
K(\rho_1)$. Using that $\rho_s\propto |\hat e,g\rangle\langle \hat
e,g\rangle$ we would have $\rho|e,h\rangle=0$ and $\rho|\hat e,h\rangle=
c |\hat e,g\rangle$ where $c$ is a constant. If we define
$|f'\rangle=c|f\rangle-|h\rangle$ we have that $|e,f'\rangle,|\hat
e,f'\rangle\in K(\rho)$, contrary to our assumption.$\Box$

Now, we turn to the case $\rho=\rho^{T_A}$, and therefore specialize
our previous results to this case. In particular, we formulate a
resut similar to Lemma 4.

{\bf Lemma 4b:} If $\rho=\rho^{T_A}$ and $\exists |e,f\rangle \in
K(\rho)$ then we can write
$\rho=\rho_{N-1} + \rho_s$, where $\rho_s$ is separable, $\rho_{N-1}$
is supported on $\C^2\otimes \C^{N-1}$ and $\rho_{N-1}^{T_A}=\rho_{N-1}$.

{\em Proof:} First, according to Lemma 2 we have that
$|e^\ast,f\rangle\in K(\rho^{T_A})=K(\rho)$ and therefore we obtain that
$|e_{\rm r},f\rangle\in K(\rho)$ with $|e_{\rm r}\rangle=|e\rangle+|e^\ast\rangle$ 
real (if this vector is zero we can
take $|e_{\rm r}\rangle=(|e\rangle-|e^\ast\rangle)/i$). Now, using the
vector $|e_{\rm r}\rangle$ instead of $|e\rangle$ we can derive the results
of Lemmas 3 and 4a but with real vectors $|\hat e_r\rangle$. 
Thus, according to 
(\ref{rho1}) we have $\rho_{1}^{T_A} =\rho^{T_A}-(\langle g|f\rangle)^{-1}
|\hat e_{\rm r}^\ast,g\rangle\langle \hat e_{\rm r}^\ast,g|=\rho_1$.$\Box$

We are not at the position of proving the main result of this paper:

{\bf Theorem 2:} If $\rho=\rho^{T_A}$ then $\rho$ is separable. 

{\em Proof:} We prove it by induction. First, in the case $2\times 1$
(i.e. $N=1$) it is obviously true. Now, let us assume that it is valid
for the case $2\times (N-1)$ and let us prove that then it is also valid
for the case $2\times N$. In order to do that, we will show that any
density operator $\rho$ can be decomposed as $\rho=\rho_1 + \rho_s$
where $\rho_s$ is separable and $\rho_1=\rho_1^{T_A}$ is supported on
$\C^2\otimes \C^{N-1}$ and therefore is also separable according to our
assumption. We consider two cases: (i) $r(\rho)\le N$: we have
dim$[K(\rho)]\ge N$; using Lemma 5, there is a product vector in
$K(\rho)$, so that according to Lemma 4b we have obtained the desired
decomposition. (ii) $r(\rho)>N$: according to Lemma 5 there is a product
vector $|e_{\rm r},g\rangle$ with $|e_{\rm r}\rangle=|e_{\rm
r}^\ast\rangle$. Using Corollary 1 we can use this product vector to
reduce the rank of $\rho$. But since $|e_{\rm r},g\rangle =|e_{\rm
r}^\ast,g\rangle$ we have that the resulting operator is also equal to
its partial transpose. We can proceed in this way until $r(\rho)=N$
which corresponds to the case (i), and therefore we complete the
proof.$\Box$

Note that we could have chosen a different basis for the partial
transposition. Taking into account that for any symmetric unitary
operator we can always find a basis in which partial transposition is
related to the original one by that such operator, we have

{\bf Corollary 2:} If $\rho=(U_A\otimes 1)
\rho^T (U_A\otimes 1)^\dagger$ for some
unitary operator $U_A=U_A^{T_A}$ then $\rho$ is separable \cite{gener}. 

% --------------- sufficient separability criterion --------------

Theorem 2 suggests that if $\rho$ is not very different from
$\rho^{T_2}$, then it should also be separable. Indeed, one can
construct a powerful sufficient separability condition based on that
theorem. We will first introduce a lemma that generalizes Lemma 1
and that is useful to determine whether the difference of two positive
operators is positive. We will use the operator norm which is
defined as usual $\|A\| = \max \| A|x\rangle\|$ with $\||x\rangle=1$.

{\bf Lemma 6:} Given two hermintian operators $X,Y\ge 0$, 
$X-Y\ge 0$ iff $Y$ 
is supported on $R(X)$ and  $\|Y^{1/2} X^{-1/2} \|^2 \le 1$. 

{\em Proof:} If $Y$ is not supported on $R(X)$ we have that $\exists
|\phi\rangle \in R(Y),K(X)$ and therefore
$\langle\phi|X-Y|\phi\rangle=-\langle\phi|Y|\phi\rangle<0$. On the
contrary, if $Y$ is supported on $R(X)$ we have that $X-Y\ge 0$ iff
$\forall |\varphi\rangle\in R(X)$ we have
$\langle\varphi|Y|\varphi\rangle/\langle\varphi|X|\varphi\rangle\le 1$.
Denoting by $|\psi\rangle=X^{1/2}|\varphi\rangle$ we obtain that $X-Y\ge
0$ iff
\be
1\ge \max_{|\psi\rangle} \frac{\langle \psi|X^{-1/2}YX^{-1/2}|\psi\rangle}
{\langle \psi|\psi\rangle} =  \|Y^{1/2}X^{-1/2}\|^2.\Box
\ee

In order to give the sufficient condition for separability, we have to
introduce some definitions. In $2 \times N$ we can always write $\rho$ as
\be
\label{dec2}
\rho=\frac{\rho+\rho^{T_A}}{2} + \frac{\rho-\rho^{T_A}}{2}
\equiv \rho_s + \sigma^A_y\otimes B,
\ee
where $2\rho_s=\rho+\rho^{T_A}$, $\sigma^A_y=i(|0\rangle_A\langle
1|-|1\rangle_A\langle 0|)$, and $2B=2B^\dagger={\rm
tr_A}[\sigma_y^A(\rho-\rho^{T_A})]$. This operator $B$ can be decomposed
as 
\be
B=\sum_{i=1}^K \lambda_i |v_i\rangle\langle v_i|.
\ee
In particular, one of such is the spectral decomposition. Given one of such
decomposition $\{\lambda_i,|v_i\rangle\}_{i=1}^K$ and a set of real numbers 
$\{a_i\}_{i=1}^K$ we define the operator
\be
C(a,\lambda,v)\equiv \sum_{i=1}^K |\lambda_i| (a_i^2|0\rangle\langle 0|
+a_i^{-2}|1\rangle\langle 1|)\otimes |v_i\rangle\langle v_i|,
\ee
which is obviously positive. We have: 

{\bf Theorem 3:} Given a decomposition of $B$
$\{\lambda_i,|v_i\rangle\}_{i=1}^K$ and a set of real numbers
$\{a_i\}_{i=1}^K$, if $\|C^{1/2}(a,\lambda,v)\rho_s^{-1/2}\|^2\le 1$,
then $\rho$ is separable.

{\em Proof:} We define $\tilde\rho_s=\rho_s-C(a,\lambda,v)=
\tilde\rho_s^{T_A}\ge 0$ according to Lemma 6. Using Theorem 2 we have
that $\tilde\rho_s$ is separable. Let $|w_i\rangle=a_i|0\rangle
-ia_i^{-1} {\rm sign}(\lambda_i)|1\rangle$. Then, it is easy to check
that
\be
\label{end}
\rho=\tilde\rho_s + \sum_{i=1}^K|\lambda_i| |w_i,v_i\rangle\langle w_i,v_i|.
\ee
which shows that $\rho$ is separable.$\Box$

Thus, we can show that a density operator is positive if we can find a
decomposition of $B$ and a set of real numbers that fulfill certain
conditions. In particular we can take the spectral decomposition of $B$
and $a_i=1$. Using the fact that $\|AB\| \le \|A\| \|B\|$ one can
easily prove the following:

{\bf Corollary 3:} If $\rho+\rho^{T_A}$ is of full range and $\|(\rho +
\rho^{T_A})^{-1}\| \|\rho-\rho^{T_A}\| \le 1$, then $\rho$ is separable.
This corollary implies that if $\rho$ is full range and is very close to
$\rho^{T_A}$ then it is separable.

% --------------------------- 2 x 2 ------------------------------
The results introduced in the first part of this Letter also allow to
prove Peres criterion \cite{Pe96} for separability in the $2\times 2$
case. In particular, in the following we will show that in these systems
if $\rho^{T_A}\ge 0$ then $\rho$ is separable. Note that the converse
can be easily proved \cite{Pe96}.

{\bf Theorem 4:} (Peres, \cite{Pe96}). If $\rho,\rho^{T_A}\ge 0$ are operators
supported on $\C^2\otimes \C^2$, then $\rho$ is separable.

{\em Proof:}
We just have to consider the case ${\rm dim}[K(\rho)]={\rm
dim}[K(\rho^{T_A})]=1$ in which both $K(\rho)$ and $K(\rho^{T_A})$
contain no product vector. The reason is that: (i) if $K(\rho)$ or
$K(\rho^{T_A})$ contain a product vector then using Lemma 4a we can
reduce the problem to the $2\times 1$ case, and therefore $\rho$ is
separable; (ii) If ${\rm dim}[K(\rho)]=2$ then according to Lemma 5
there is a product vector in $K(\rho)$, so that we are back in (i) and
therefore $\rho$ is separable; (iii) similarly, if ${\rm
dim}[K(\rho^{T_A})]=2$ we conclude that $\rho^{T_A}$ and therefore
$\rho$ are separable; (iv) If ${\dim K}(\rho^{T_A})=0$ and ${\dim
K}(\rho)<2$ then according to Lemma 5 $R(\rho)$ contains at least one
product vector $|e,f\rangle$ and obviously $|e^\ast,f\rangle\in
R(\rho^{T_A})$; we can apply Corolary 1 and substract product vectors
from $\rho$ until either ${\rm dim}K(\rho)=2$, in which case $\rho$ is
separable, or ${\rm dim}[K(\rho)]={\rm dim}[K(\rho^{T_A})]=1$; (v)
Similarly If ${\dim K}(\rho)=0$ and ${\dim K}(\rho^{T_A})<2$ we arrive
at the same conclusion. Thus, let us assume that $\rho|\Psi_1\rangle=0$
and $\rho^{T_A}|\Psi_2\rangle=0$, where $|\Psi_{1,2}\rangle$ are of the
form (\ref{Psi}) and they are not product vectors. We will first show
that these vectors can always be writen as
\bma
\label{dec}
\bea
|\Psi_1\rangle &\propto& |e,f\rangle - |\hat e,g\rangle,\\
|\Psi_2\rangle &\propto& |e^\ast,h\rangle - |\hat e^\ast,f\rangle,
\eea
\ema
where both $\{|f\rangle,|g\rangle\}$ ($\{|f\rangle,|h\rangle\}$)
are linearly independent since otherwise $|\Psi_1\rangle$ ($|\Psi_2\rangle$)
would be a product vector. Then we will show that either $\rho=\rho^{T_A}$
or $|h\rangle=k^2|g\rangle$ where $k^2>0$. In the first case, according to
Theorem 2, $\rho$ is separable. In the second case, we have that if we define
$|a\rangle=|e\rangle + k |\hat e\rangle$ and $|b\rangle = \langle f|\hat g\rangle
|\hat f\rangle - k \langle g|\hat g\rangle |\hat g\rangle$ then
the vector $|a,b\rangle\in R(\rho)$ and $|a^\ast,b\rangle\in R(\rho^{T_A})$, since
they are orthogonal to $|\Psi_{1,2}\rangle$ respectively. Thus, according
to Corolary 1 we can substract this product vector and increase the
dimension of either $K(\rho)$ or $K(\rho^{T_A})$ to 2, which as we showed
in (ii) and (iii) implies that $\rho$ is separable.

We show how to obtain the decomposition (\ref{dec}). We look for two
unnormalized vector $|e\rangle = \alpha|0\rangle + |1\rangle$ and
$|\tilde f\rangle=f_0|0\rangle+f_1|1\rangle$ such that $\langle
e,\tilde f|\Psi_1\rangle = \langle \hat e^\ast,\tilde f|\Psi_2\rangle=0$. We obtain
two linear equations for $f_0$ and $f_1$ of the form $C(\alpha) \vec
f=0$, where $C$ is a matrix that depends linearly on $\alpha$ (and not
on $\alpha^\ast$ \cite{note}). These equations always have a solution
for some given $\alpha$ since the condition ${\rm det}[C(\alpha)]=0$
is a second order equation for $\alpha$. 

We finally show that either $\rho=\rho^{T_A}$ or
$|h\rangle=k^2|g\rangle$ where $k^2>0$. First, we show that 
\be
\label{state}
\langle f|\rho|f\rangle = \langle g|\rho|h\rangle = \langle h|\rho|g\rangle.
\ee
To obtain that we use the fact that the vectors $|\Psi\rangle$ are in
the kernel of $\rho$ or $\rho^{T_A}$, and therefore
$\rho|e,f\rangle=\rho|\hat e,g\rangle$ and
$\rho^{T_A}|e^\ast,h\rangle=\rho^{T_a}|\hat e^\ast,f\rangle$. Using
(\ref{property}) we can write the second equation as $\langle
e|\rho|h\rangle = \langle \hat e|\rho|f\rangle$. Using these equations
it is easy to obtain that $\langle e,f|\rho|f\rangle=\langle e,g|\rho|h\rangle$ 
and $\langle \hat e,f|\rho|f\rangle=\langle \hat
e,g|\rho|h\rangle$ which automatically proves the statement (\ref{state}).
Now, given that $|f\rangle$ and $|g\rangle$ are linearly independent
vectors, we can write $|h\rangle=\alpha|f\rangle+\beta|g\rangle$ with
$\beta\ne 0$. If $\alpha=0$, then using (\ref{state}) we have that
$\beta\equiv k^2>0$ is real and positive. If $\alpha\ne 0$ we have
that both $\langle g|\rho|f\rangle$ and $\langle f|\rho|g\rangle$
can be expressed as linear combinations of $\langle f|\rho|f\rangle$
and $\langle g|\rho|g\rangle$. These two last operators are hermitian and
they are equal to their transpose in some given
basis (if we write $\langle f|\rho|f\rangle\propto 1+\vec n_f\cdot \sigma$
and $\langle g|\rho|g\rangle\propto 1+\vec n_g\cdot \sigma$ then we just
have to use the basis such that both $\vec n_f$ and $\vec n_g$ lie in the
$x-z$ plane). Thus, $\langle g|\rho|f\rangle$ and $\langle f|\rho|g\rangle$
are also equal to their transposes in that basis. Since $|f\rangle$ and
$|g\rangle$ are linearly independent we immediately arrive to the conclusion
that if we consider the partial transpose in that basis 
we have that $\rho=\rho^{T_A}$, which completes the proof.$\Box$

Summarizing, we have demonstrated that all density operators on $2\times
N$ that remain invariant after partial transposition with respect to the
first system are separable. Using this fact we have constructed a
sufficient separability condition for such systems. We have also given a
relatively simple proof of Peres criterion based on these results.

This work has been supported by Deutsche Forshungsgemeinschaft (SFB
407), the \"Osterreichisher Fonds zur F\"rderung der wissenschaftlichen Forschung
(SFB F15), and the European TMR network ERB-FMRX-CT96-0087. J. I. C.
thanks the Univesity of Hannover for hospitality. We thank A. Sanpera
and G. Vidal for fruitful discussions.

% -------------------------------------------------------------


\begin{references}
\bibitem{gener}
This corolary can be generalized using arbitrary nonsingular matrices
instead of unitary ones.
\bibitem{note}
Note that if the coefficients depend on $\alpha^\ast$ as well it might happen
that the second order polynomial equation does not have solutions. This is
the reason why we cannot make use of Lemma 5 directly to show that there
exists a vector $|a,b\rangle\in R(\rho)$ and $|a^\ast,b\rangle\in R(\rho^{T_A})$.
\end{references}
\end{document}